\documentclass[useAMS,usenatbib]{mn2e}
\usepackage{epsfig}

\title[The M Dwarf Problem in the Galaxy]{The M Dwarf Problem in the Galaxy}
\author[Vincent M. Woolf and Andrew A. West]{Vincent M. Woolf$^{1}$\thanks{
E-mail: vwoolf@unomaha.edu; aawest@bu.edu} and Andrew A. West$^{2}$
\\
$^{1}$University of Nebraska at Omaha, Physics Department, 6001 Dodge Street,
Omaha, NE 68182, USA \\
$^{2}$Boston University, Department of Astronomy, 725 Commonwealth Avenue,
Boston, MA 02215, USA}
\begin{document}

\date{}


\maketitle

\label{firstpage}

\begin{abstract}
We present evidence that there is an M dwarf problem similar to the previously
identified G dwarf and K dwarf problems: the number of low-metallicity M dwarfs
is not sufficient to match simple closed-box models of local Galactic
chemical evolution.  We estimated the metallicity of 4141 M dwarf stars with
spectra from the Sloan Digital Sky Survey (SDSS) using a molecular band strength
versus metallicity calibration developed using high resolution spectra of
nearby M dwarfs.  Using a sample of M dwarfs with measured magnitudes,
parallaxes, and metallicities, we derived a relation that describes the absolute
magnitude variation as a function of metallicity.  When we examined the
metallicity distribution of SDSS stars, after correcting for the different
volumes sampled by the magnitude-limited survey, we found that there is an M
dwarf problem, with the number of M dwarfs at $\rm [Fe/H] \sim -0.5$ less
than 1\% the number at $\rm [Fe/H] = 0$, where a simple model of Galactic
chemical evolution predicts a more gradual drop in star numbers with decreasing
metallicity.

\end{abstract}


\section{Introduction}
The compositions of stars provide a test for models of Galactic chemical
evolution, with the general assumption being that the photospheric chemical
composition of most stars represents the local Galactic chemical composition 
where they formed.
van den Bergh \citeyearpar{v62} and Schmidt \citeyearpar{s63}
first noted that the ratio of
low-metallicity G dwarfs to solar-metallicity G dwarfs in the local 
neighbourhood is too small to be explained by the `Simple model' of
Galactic chemical evolution. The Simple
model, as summarized by Tinsley \citeyearpar{t80}, is a model assuming (1) the
solar neighbourhood can be modeled as a closed system; (2) it started as 100\%
metal-free gas; (3) the initial stellar mass function (IMF) is constant;
and (4) the gas is chemically homogeneous at all times. Each of the
assumptions in the Simple model are demonstrably false, but it is an
important starting point for the development of more complex
models.  The process of solving the G dwarf problem and explaining
the Galaxy's chemical enrichment history 
has been underway for nearly five decades. The solution will
require discarding one or more the Simple model's assumptions. The
introduction of a variable IMF \cite[e.g.][]{s63, c96, m00, r05}, variable
star formation rates with intermittent mixing \cite[e.g.][]{m93, c08}, and / or
inflow or outflow of material \cite[e.g.][]{w95, p01} into models of Galactic
chemical evolution produces metallicity distributions in better agreement
with observations.

The G dwarf problem does not apply only to the local neighbourhood: the
G dwarf problem exists in other galaxies as well \citep*{w96}. While G dwarf
lifetimes would suggest that some G dwarfs may have left the main
sequence during the Galaxy's
lifetime, the paucity of low-metallicity stars locally also extends to the
longer-lived K dwarfs \citep{c04}. Mould \citeyearpar{m78a} suggested that there
may be an M dwarf problem: using spectra of six M
dwarfs and infrared spectroscopy of sixteen old disc M dwarfs, he
estimated the abundance dispersion for M dwarfs and found evidence that
suggested a common chemical history of G and M dwarfs.

We have directly tested the existence of an M dwarf problem by
estimating the metallicity of 4141 M dwarf stars with temperatures in 
the range where our analysis method is valid,
$\rm 3500 K \le T_{eff} \le 4000 K$, using CaH and TiO molecular band
strengths measured from Sloan Digital Sky Survey (SDSS) spectra
\citep{y00}  and
find that there is an M dwarf problem in the Galactic disc similar to the
previously known G and K dwarf problems.  This does not suggest a new
solution to the M, K, and G dwarf problems, but rather shows that for all
stars with main sequence lifetimes comparable or larger than the age of the
Galaxy, and where chemical abundance surveys have been performed, the number
of low-metallicity stars is insufficient to match the Simple model.

\section{Estimating metallicities}
M dwarfs are inherently faint objects, and therefore
must be nearby to appear bright
enough to obtain high-resolution spectra,
$\lambda / \Delta \lambda \ga 30,000$, of sufficient
quality to measure atomic absorption lines for abundance analyses, while using
a reasonable amount of telescope time. 

Molecular band indices CaH2, CaH3, and TiO5 \citep*{r95} measured from
low-resolution spectra, $\lambda / \Delta \lambda \sim 1800$, have been shown
to be useful in classifying cool dwarfs. CaH2 and CaH3 correlate well with
spectral type and the combination of CaH2 or CaH3 with TiO5 separates cool
dwarfs into rough metallicity classes: dwarfs,
subdwarfs, and extreme subdwarfs \citep{g97, l03}. 
Woolf, L\'epine, \&
Wallerstein \citeyearpar{w09} used metallicities of 88 M dwarfs, estimated using
atomic lines measured from high-resolution spectra \citep{w05,w06}, to develop
and calibrate a method to use CaH2, CaH3, and TiO5 molecular band indices
to estimate M dwarf metallicities more precisely.  CaH2, CaH3, and their sum
(CaH2 + CaH3) were found to correlate well with effective temperature
($\rm T_{eff}$) and TiO5 was found to vary with
both $\rm T_{eff}$ and metallicity \citep{w06}. The three indices were
used with the calculated [Fe/H]\footnote{where 
$\rm [X] = \log_{10}(X)_{star} - \log_{10}(X)_{Sun}$.}
metallicites of the 88 M dwarfs to derive the
relation
\begin{equation}
{\rm [Fe/H]} = a + b (\zeta_{TiO/CaH}),
\end{equation}
where $a = -1.685 \pm 0.079$, $b = 1.632 \pm 0.096$, and the metallicity
index $\rm \zeta_{TiO/CaH}$
is defined as described in L\'epine, Rich, \& Shara \citeyearpar{l07}.
This calibration of the molecular band index versus metallicity relation
is valid for stars with
$\rm 3500 K \le T_{eff} \le 4000 K$ and $\rm -1.5 \le [Fe/H] \le +0.05$.
As described in \citet{w09}, the [Fe/H] versus $\rm \zeta_{TiO/CaH}$
fit, should be accurate to $\pm 0.3$ dex for stars with $\rm [Fe/H] > -1.0$ and
to $\pm 0.5$ dex for stars with $\rm [Fe/H] < -1.0$.
based on the maximum errors the empirical data appear to allow, as opposed
to standard statistical errors based on repeated measurements of the same
quantity.  Most errors should be less than half of those maxima.

We used CaH2, CaH3, and TiO5 indices measured from SDSS spectra to 
estimate metallicities for the 4141 M dwarfs in our sample.  The stars were
placed in 0.1 dex [Fe/H] bins from $-$1.5 to 0.1. This extrapolates
slightly beyond the limits for which our method was calibrated. 
The number of stars in bins with [Fe/H] less than $-$0.50 is very
small, so extrapolating to [Fe/H] = $-$1.55 at the low metallicity end makes
little difference. The number of stars in the $\rm [Fe/H] \approx +0.1$ bin is
less certain because the entire bin is outside our calibration, but the bin is
included to show that, if our calibration continues to be correct a little
beyond where testing stopped because of a lack of stars for the calculation,
it appears that the trend
in number of stars versus metallicity drops for $\rm [Fe/H] > 0$.

\section{Stellar data set}

Our spectroscopic sample contains 4141 low-mass stars selected from the SDSS
\citep{y00} Munn `special plates,' which were
made public as part of the SDSS Data Release 4 \citep{a06}.
The Munn plates were targeted to be a magnitude-limited sample of M and late-K
dwarfs with $i< 18.26$, $i-z > 0.2$ and be located near the $l = 123^\circ$,
$b=-63^\circ$
Galactic sight line.  This sample was previously used to study the kinematics
of the local Milky Way thin and thick discs \citep{b07a} to roughly 2 kpc
above the Galactic plane and are
contained in the larger SDSS M dwarf spectroscopic samples \citep{w04,w08,w11}.
We analysed the 8859 spectra that were
observed using the Hammer spectral typing facility \citep{c07} and
rejected stars that had poor quality (SNR $<$ 3 near H$\alpha$) and spectral
types earlier than K5 (resulting in 7714 stars). 

As part of our analysis, we computed radial velocities (RV) by
cross-correlating each spectrum with the appropriate \citep{b07b}
M dwarf template.  In the case of K5 and K7 dwarfs, we used the M0 template for
our RV determinations.  The computed velocities were used to correct the 
stellar spectra to zero radial velocity.  We then computed the TiO5, CaH2 and
CaH3 molecular band indices \citep{r95, g97} from the velocity-corrected
spectra. Additional stars were removed from the sample if their molecular
band indices, temperature, or [Fe/H] fell outside the range covered by the
Woolf et al. \citeyearpar{w09} metallicity calibration, leaving the 4141
stars for our study.  All stars with
temperature and [Fe/H] within the metallicity calibration have $i-z$ colours
larger the 0.2 value used in the initial selection, so the colour limit
should introduce no selection effects for our metallicity statistics.

\section{Volume and stellar number density corrections}
\subsection {Volume correction for $V$ and $R$ magnitudes}
Low metallicity main sequence stars (subdwarfs) are less luminous than higher
metallicity stars with the same temperature or spectral type
\citep{r54, mm78, a80, b11}.  Because the
SDSS M dwarf sample was selected to fall within a magnitude range, the
subdwarf M stars must be closer on average than the higher metallicity stars,
and thus sample a smaller volume of our Galactic neighbourhood.
We must correct for this to avoid introducing a Malmquist Bias-like effect and
overcounting the number of more luminous stars.

\begin{figure}
\epsfig{file=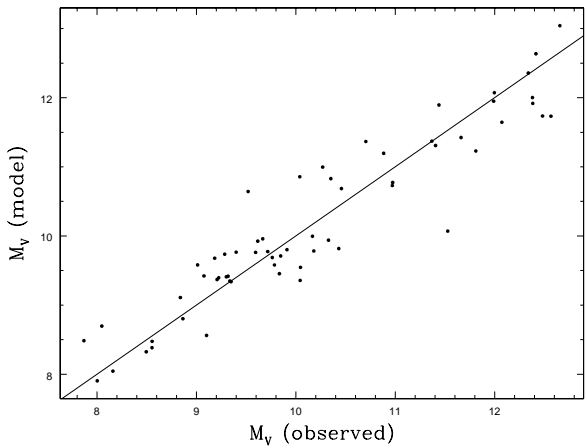, width=6.2cm, angle=270}
\caption{Absolute V magnitudes for early M dwarfs calculated using Equation
\ref{eq:eq2} versus those calculated using observed V magnitudes and parallax
measurements.  The diagonal line indicates where points would fall if the model
and observed values were identical.}
\label{F1}
\end{figure}

We used stars from \citet{w09} for which parallax data were available and thus
absolute magnitudes could be calculated to find the luminosity variation with
metallicity for M dwarfs in our temperature range. An absolute
magnitude difference
function was found by fitting a three dimensional `surface' to the [Fe/H],
CaH2$+$CaH3, and $\rm M_V$ data points calculated for the stars.
CaH2$+$CaH3 acts as a temperature proxy in the calculated fit.  We found that
for the best-fitting surface, the magnitude change with metallicity
did not depend on CaH2$+$CaH3 (temperature)
within the range of our metallicity calibration. For example, two 3500 K stars
with [Fe/H] = 0.0 and -1.0 differ in magnitude by the same amount as two 4000 K
stars with [Fe/H] = 0.0 and -1.0. The fit to the data points is
\begin{eqnarray}
\label{eq:eq2}
\rm{M_V} = a + b\rm{[Fe/H]} + c\rm{[Fe/H]}^2 + d(\rm{CaH2 + CaH3}) \\
 + e(\rm{CaH2 + CaH3})^2, \nonumber
\end{eqnarray}
where $a= 25.356$, $b= -0.6394$, $c= 0.8455$, $d= -15.778$, and
$e= 3.250$. The root mean squared deviation between the absolute V magnitudes
calculated using parallax and apparent V magnitudes and those calculated using
the above formula is 0.450.
The $\rm M_V$ values calculated from Equation \ref{eq:eq2}, $\rm M_V$ (model),
are compared to those calculated from observed $V$ magnitudes and trignometric
paralax, $\rm M_V$ (observed), in Figure \ref{F1}.
We did a similar fit using the absolute R magnitudes and found coefficients
$a= 24.334$, $b= -0.8258$, $c= 0.6964$, $d= -16.379$, and $e= 3.696$, with
the root mean squared deviation between the $\rm M_R$ values calculated from
paralax and R measurements and from the formula being 0.485.

Because there is no temperature dependence in the difference in magnitude
(or luminosity) for M dwarfs of the same temperature but with different
metallicities there is also no temperature dependence in the volume correction
due to metallicity.  We can apply the same volume sampling correction for all
stars in our sample without considering temperature.

The magnitude-limited volume sampled for each metallicity bin
is proportional to the distance cubed. The distance to a star is
proportional to the square root of its luminosity for a given
apparent magnitude.
So for a given stellar luminosity, L, the sampled volume is proportional
to $ L^{3/2}$.  Subdwarfs with smaller luminosities are sampled from a smaller
volume, so we multiply
the numbers of stars in the metallicity bins by the inverse of the volume
factor, $L^{-3/2}$, to correct for the volume difference.
The $V$ and $R$ luminosity and absolute magnitude differences and the volume
correction factor for different [Fe/H] values as calculated using \citet{w09}
stars are shown in Table \ref{T1}. 

\subsection{Volume correction for SDSS $i$ magnitude}
The SDSS stars in our sample were selected for observation using an SDSS $i$
magnitude cutoff. We must therefore use $\Delta M_i$, the difference in 
$i$ absolute magnitude caused by varying metallicity, in our volume correction.

None of the stars used for our volume correction due to
metallicity differences have SDSS photometry available, so they cannot be
used to calculate $\Delta M_i$ directly.
Instead, we first use $\Delta M_V$ and $\Delta M_R$  to
estimate $\Delta M_r$, the difference in SDSS $r$ magnitude due to metallicity
differences. We then use $\Delta (r-i)$, the difference in $r-i$ colour due
to metallicity differences, to calculate $\Delta M_i$ to use for our volume
correction.

The SDSS $r$ photometric band overlaps the $V$ and $R$ bands.  Differences in
metallicity should therefore cause differences in $r$ absolute magnitude similar
to the differences found for $V$ and $R$. The corrections for  $V$ and $R$
differ at most by 0.06 magnitudes.  Assuming that $\Delta M_r$ is equal to the
mean of $\Delta M_V$ and $\Delta M_R$ should therefore introduce errors smaller
than a few percent.

We have therefore estimated the necessary
corrections using updated synthetic spectra (Hauschildt, private communication)
based on PHOENIX calculations \citep{h99}, finding the mean differences in
$r - i$ colours in the spectra relative to the [Fe/H] = 0.0 spectrum for
temperatures between 3500 K and 4000 K. We then calculated
the difference in $i$ magnitude due to metallicity differences:
$\Delta M_i = \Delta M_r - \Delta (r-i)$. The corresponding $i$ luminosity
ratio and volume corrections were calculated in the manner described
previously for $V$ and $R$ and are reported in Table \ref{T1}.

An alternate method to calculate $\Delta (r-i)$ would use a published
empirical colour-temperature calibration to estimage $B-V$, $V-R$, and $V-I$
colours for stars of M dwarf temperatures then use an existing
empirically-derived formula to transform these to an $r-i$ colour. 
\citet{wl11} provide such a colour-temperature calibration. The M dwarf
metallicities used in their calibration are based on kinematics, with young
disc stars assigned $\rm [Fe/H] = -0.1$, old disc stars $-0.5$,
and halo stars $-1.5$, unless the star is in a cluster, in
which case the cluster metallicity was used.  If we use their empirical
calibration rather than the colours from synthetic spectra, we find much
smaller corrections for $\Delta (r-i)$.  For example, the  $r-i$ correction
for $\rm [Fe/H] = -1.5$ stars is $-0.3$ if calculated from synthetic spectra,
but is about 0.02 if calculated from the empirical calibration data.

Our goal is to compare the trend for M dwarf stars with a model which
has predicted more low-metallicity stars than have been observed. We will
therefore avoid the possibility of under-counting low-metallicity stars by
using the larger correction based on synthetic spectra which tends to
increase the corrected number of low-metallicity stars.  In the end, because the
number of observed stars drops off so quickly with decreasing metallicity,
our conclusions do not depend on which of these corrections we use. We
correct by multiplying the number of stars in each metallicity bin by the values
in the $\rm volume^{-1} (i)$ column in Table \ref{T1}.  The empirically-based
difference between $\Delta M_V$ and $\Delta M_i$ is very small, so if someone
prefers not to use the correction based on synthetic spectra, one can multiply
by the value in the $\rm volume^{-1} (V)$ column instead.

We note that the SDSS sample has both a faint $i$ magnitude limit and a
bright $i$ magnitude limit, meaning that the volumes sampled are effectivly
spherical shells.  The ratio of the distances to the faint magnitude cutoff
is the same as the ratio of the distances to the bright magnitude cutoff
however, so the volume ratio of the unsampled inner spheres is the same as the
volume ratio of the spheres defined by the faint limit: no additional
correction is required to account for the bright magnitude cutoff.

\subsection{Correction for stellar number density differences}
The observational direction for our
SDSS M dwarf sample, near $l = 123^\circ$, $b=-63^\circ$, means that 
the more distant stars in the sample are farther from the plane of the
Galactic disc and thus in a region with a lower stellar number density.
Without correcting for this effect we would
overestimate the fraction of fainter,
low-metallicity subdwarf stars which are found in a denser region of the disc.
We use the difference in distances due to
metallicity and a model of the Milky Way stellar number density distribution
to correct for this effect.

\citet{b11} report absolute $r$ magnitudes, $M_r$, for M dwarfs based
on photometric parallax. Figure~2 from their paper plots $M_r$ vs spectral
type. From this figure we find that the typical $M_r$ for stars
in our data set, mostly M0 to M2 dwarfs, is about 9.0. This is
consistent with the absolute magnitude of the few warmer M dwarfs 
with reported SDSS photometry and trigonometric parallax \citep{d06}.
A $M_r = 9.0$ star with $r = 17.5$,
the mean observed $r$ magnitude for our sample, is at a distance $d = 501$ pc.

\citet{j08} report the stellar number density distribution for
the Milky Way, providing an exponential equation to calculate
estimates of changes in density
with radial distance from Galactic centre and perpendicular distance from the
plane of the disc (Juric et al. Equation 23).  Stellar number density
variation due to differences in radial distance from Galactic centre is minimal
for our stars, given the direction and distance to the sample.
We assume the radial terms from the Juric et al. equation are constant
and recognize that $Z_\odot$ is above the plane of the Galaxy and the direction
to the stars in our sample places them below the plane. The 
equation thus becomes
$\rho = C \exp(\frac{Z_\odot - Z}{H}$),
where C is the stellar number density
at the vertical centre of the disc at the Solar radial distance, $Z_\odot$ is
the Solar distance above the Galactic plane, $Z$ is the distance
to the star measured in the direction perpendicular below the plane, and
$H$ is the scale height for the disc, thin or thick.  Note that this form of
the equation is not valid for stars above the plane or closer
to the centre of the plane than the Sun.  We use this modified
equation to find the ratio of the stellar number density $\rho$ for a star in
our sample with a distance $Z$ compared to the local density $\rho_\odot$:
\begin{equation}
\label{eq:density}  
\frac{\rho}{\rho_\odot} = \exp(\frac{ - Z}{H}).
\end{equation}
\cite{j08} report $H= 300$ pc for the thin disc, $H = 900$ pc for the
thick disc, and  $Z_\odot = 20 $ pc.

To make a rough correction for differences in stellar number density caused by
metallicity we first use $\Delta M_r$ from Table \ref{T1} to find the
luminosity ratio relative to a [Fe/H] = 0 star for each metallicity bin:
$l / l_0 = 100^{- \Delta M_r / 5}$.  The ratio of the distances for stars with
the same apparent magnitude is then
$d / d_0 = \sqrt{l / l_0}$. If we then assume that the mean distance to a
[Fe/H] = 0 star in our sample is 501 pc we can use the distance ratio to find
the mean distance to stars in the other metallicity bins.  Because
$b \approx -63^\circ$ for our sample, the distance perpendicular to the Galactic
plane is $Z = d \sin 61^\circ - Z_\odot$.

We calculate the ratio of the stellar number density at the typical
perpendicular distance, $Z$, for stars in each metallicity bin relative to the
local Solar neighborhood density using
Equation \ref{eq:density}.  We calculate the ratio for both thin disc stars,
$(\frac{\rho}{\rho_\odot})_{\rm thin}$, and thick disc stars,
$(\frac{\rho}{\rho_\odot})_{\rm thick}$, and use these and the local
12\% thick disc fraction \citep{j08} to calculate the density ratio 
$\frac{\rho}{\rho_\odot} = 0.88 (\frac{\rho}{\rho_\odot})_{\rm thin} + 0.12
(\frac{\rho}{\rho_\odot})_{\rm thick}$.  For each metallicity bin we divide
this density ratio $\frac{\rho}{\rho_\odot}$ by the value found for the
[Fe/H] = 0 bin to get $\frac{\rho}{\rho_0}$, where ${\rho_0}$ is the
stellar number density at the typical distance to solar metallicity stars in
our sample.  To correct for differences in stellar number density
we multiply the number of stars in each
metallicity bin by the inverse of this density ratio,
$(\frac{\rho}{\rho_0})^{-1}$, as reported in Table \ref{T1}.
We note that the volume correction and the stellar number density correction
act in opposite directions: fainter, lower metallicity stars are sampled in
a smaller volume, but in a region with a higher density of stars.

\begin{table*}
 \centering
 \begin{minipage}{140mm}
  \caption{M dwarf absolute magnitude differences and luminosity,
volume, and stellar number density
correction factors for sampling differences compared to Solar [Fe/H]. Values in
the $\rm volume^{-1}(i)$ and $(\frac{\rho}{\rho_0})^{-1}$
columns are the factors by which the numbers of stars
in each [Fe/H] bin should be multiplied to correct for volume sampling
and stellar number density
differences. $\Delta (r - i)$ is the colour difference relative to [Fe/H] = 0.0
due to metallicity difference.}\label{T1}
  \begin{tabular}{@{}cccccccccccc@{}}
  \hline
   [Fe/H]& $\rm \Delta M_V$  & $\rm \Delta M_R$& L($V$)& 
    L($R$)& $\rm volume^{-1}(V)$ & $\Delta M_r$ &
   $\Delta (r - i)$ & $\Delta M_i$ &
   L($i$) & ${\rm volume^{-1}}(i)$ & $(\frac{\rho}{\rho_0})^{-1}$\\
 \hline
$+0.1$ & $-$0.056 & $-$0.075 & 1.056 &1.072 &  0.925 &$-$0.066& $+ 0.050$ &$-$0.116&1.112&0.852 &1.038\\
 0.0 &    0.000 &    0.000 & 1.000 &1.000 &  1.00  &0.00& 0.000 &0.00&1.00& 1.00&1.00\\
$-0.1$ &    0.072 &    0.090 & 0.936 &0.920 &  1.10  &0.081& $-0.037$& 0.118&0.897&1.18 &0.957\\
$-0.2$ &    0.161 &    0.193 & 0.862 &0.837 &  1.25  &0.177& $-0.074$ &0.251&0.794&1.41 &0.909\\
$-0.3$ &    0.268 &    0.311 & 0.781 &0.751 &  1.45  &0.289& $-0.111$ &0.400&0.692&1.74 &0.859\\
$-0.4$ &    0.391 &    0.442 & 0.698 &0.656 &  1.72  &0.416& $-0.148$ &0.564&0.595&2.18 &0.807\\
$-0.5$ &    0.531 &    0.587 & 0.613 &0.582 &  2.08  &0.559& $-0.184$ &0.743&0.504&2.79  &0.756\\
$-0.6$ &    0.688 &    0.746 & 0.531 &0.503 &  2.59  &0.717& $-0.203$ &0.920&0.429&3.56  &0.707\\
$-0.7$ &    0.862 &    0.919 & 0.452 &0.429 &  3.29  &0.890& $-0.222$ &1.112&0.359&4.65  &0.659\\
$-0.8$ &    1.052 &    1.105 & 0.380 &0.361 &  4.28  &1.079& $-0.240$ &1.319&0.297&6.18  &0.615\\
$-0.9$ &    1.260 &    1.306 & 0.313 &0.300 &  5.70  &1.283& $-0.258$ &1.541&0.242&8.41  &0.573\\
$-1.0$ &    1.485 &    1.521 & 0.255 &0.246 &  7.78  &1.503& $-0.277$ &1.780&0.194&11.7  &0.535\\
$-1.1$ &    1.726 &    1.749 & 0.204 &0.200 &  10.9  &1.738& $-0.284$ &2.022&0.155&16.3  &0.501\\
$-1.2$ &    1.985 &    1.991 & 0.161 &0.160 &  15.5  &1.988& $-0.291$ &2.279&0.123&23.3  &0.470\\
$-1.3$ &    2.260 &    2.247 & 0.125 &0.126 &  22.7  &2.253& $-0.299$ &2.552&0.953&34.0  &0.442\\
$-1.4$ &    2.552 &    2.518 & 0.0953 &0.0984 & 34.0  &2.535& $-0.306$ &2.841&0.0730&50.7 &0.418\\
$-1.5$ &    2.861 &    2.802 & 0.0717 &0.0757 & 52.1  &2.831& $-0.313$ &3.144&0.0552&77.0 &0.397\\
\hline
\end{tabular}
\end{minipage}
\end{table*}

\section{M dwarf metallicity trend}
\begin{figure}
\epsfig{file=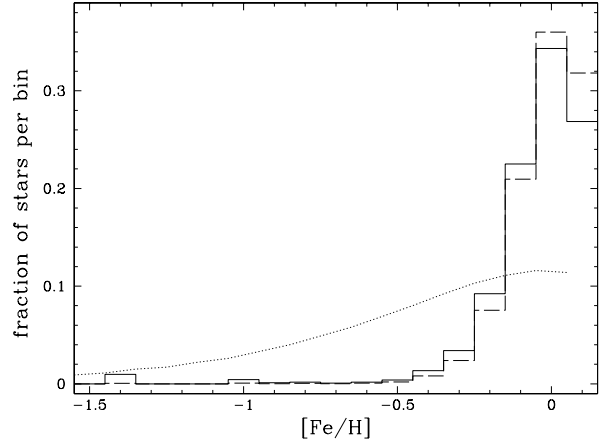, width=6.2cm, angle=270}
\caption{Fraction of stars found in each 0.1 dex [Fe/H] bin. The dashed line
represents the raw data.  The solid line represents the volume and stellar
number density corrected data.
The dotted curve is the distribution
predicted by the Simple Model.}
\label{F2}
\end{figure}

The raw and corrected fractions of M dwarfs in our sample in
each metallicity bin are shown in Table \ref{T2} and in Figure \ref{F2}.
The distribution is centred at about [Fe/H] = 0.0 and has a gaussian full
width half maximum (FWHM) of 0.30 dex and a standard deviation (dispersion)
of 0.13 dex.

\begin{table*}
 \centering
 \begin{minipage}{140mm}
  \caption{Numbers and sample fraction of M dwarfs at given metallicities.  The
last two columns are corrected for different volumes, ${\rm volume^{-1}}(i)$,
and stellar number densities, $(\frac{\rho}{\rho_0})^{-1}$,
being sampled for different metallicities.} \label{T2}
  \begin{tabular}{@{}ccccc@{}}
  \hline
   [Fe/H]& number & fraction & corrected number & corrected fraction\\
 \hline
$+0.1$ & 1318 & 0.3183 & 1166.3 & 0.2685 \\
 $0.0$ & 1491 & 0.3601 & 1491.0 & 0.3433 \\
$-0.1$ & 868  & 0.2096 & 977.7  & 0.2251 \\
$-0.2$ & 312  & 0.0753 & 401.0  & 0.0923 \\
$-0.3$ & 99   & 0.0239 & 147.9  & 0.0340 \\
$-0.4$ & 33   & 0.0080 & 58.1   & 0.0134 \\
$-0.5$ & 8    & 0.0019 & 16.9   & 0.0039 \\
$-0.6$ & 3    & 0.0007 & 7.75   & 0.0018 \\
$-0.7$ & 1    & 0.0002 & 3.07   & 0.0007 \\
$-0.8$ & 2    & 0.0005 & 7.60   & 0.0018 \\
$-0.9$ & 1    & 0.0002 & 4.82   & 0.0011 \\
$-1.0$ & 3    & 0.0007 & 18.8   & 0.0043 \\
$-1.1$ & 0    & 0      & 0      & 0      \\
$-1.2$ & 0    & 0      & 0      & 0      \\
$-1.3$ & 0    & 0      & 0      & 0      \\
$-1.4$ & 2    & 0.0005 & 42.3   & 0.0097 \\
$-1.5$ & 0    & 0      & 0      & 0      \\
\hline
\end{tabular} 
\end{minipage}
\end{table*}

Other studies have found that the number of stars peaks at a
metallicity centred in the range
$\rm -0.25 \la [Fe/H] \la 0.0$ for G and K dwarfs (e.g. \citealt{w95,
r96}; \citealt*{f97}; \citealt{j00, r00, a04, c04, n04, l05}). We find that
the peak for M dwarfs is at the high end of this range and note that our
peak is more narrow than that typically found.
There has been some ``narrowing''
of the reported metallicity distribution over time since the first reports
of the G dwarf problem in the 1960's, probably as a result of a reduction of
the uncertainties as the methods for calculating elemental abundances have
improved.

Our results for M dwarfs most closely match those found by
Favata et al. \citeyearpar{f97} for K dwarfs, however. 
In their study of G and K dwarfs,
they separated stars into two groups with effective temperatures hotter and
cooler than 5100 K and found very different metallicity distributions. They
did not report mean and standard deviations for their corrected distributions,
but we estimate from the values on their plots that the
warmer stars showed a distribution similar to those previously reported for G
dwarfs, centred at $ \rm [Fe/H] \approx -0.28$ with $\sigma \approx 0.29$, and
that the cooler star distribution dropped off more sharply toward lower
metallicities and
was more sharply peaked with a centre at $ \rm [Fe/H] \approx -0.03$ and
$\sigma \approx 0.17$.
Luck \& Heiter \citeyearpar{l05}, report elemental abundances for stars within
15 pc. Their sample contains mostly G and K dwarfs. When we separate their stars
by temperature we find that the metallicity distribution for the cool stars,
$\rm T_{eff} \le 5100 K$, has mean
$ \rm [Fe/H] = 0.0$ and  $\sigma = 0.17$, while the warmer stars have mean
$ \rm [Fe/H] = -0.11 $ and  $\sigma = 0.28$, again showing an identifiable
difference with temperature. It appears that the trend of higher mean
metallicity and narrower peaks seen for the metallicity distributions for
cooler stars seen in the Favata et al. and Luck \& Heiter data applies to
the even cooler M dwarfs in our sample.

`Simple' models of Galactic chemical evolution produce higher numbers of
long-lived low metallicity stars than we find for M dwarfs and others have
found for G and K dwarfs.  \citet{a76} use the equation
\begin{equation}
\label{eq:tinsley}
\frac{S}{S_1} = \frac{1-\mu_1^{Z/Z_1}}{1-\mu_1}
\end{equation}
to calculate ${S}/{S_1}$, the present day cumulative stellar metallicity
distribution, or the fraction of stars that have metalliciies
less than $Z$. In equation \ref{eq:tinsley},
$Z_1$ is the present day metallicity
and $\mu_1$ is the fraction of the local baryonic matter that is now 
interstellar matter (as opposed to that contained in stars and stellar
remnants). 
The smooth dotted curve in Figure \ref{F2} is derived from the
cumulative distribution described by equation \ref{eq:tinsley}, given that the
increase in the cumulative distribution must be caused by stars in each
successive metallicity bin.

\begin{figure}
\epsfig{file=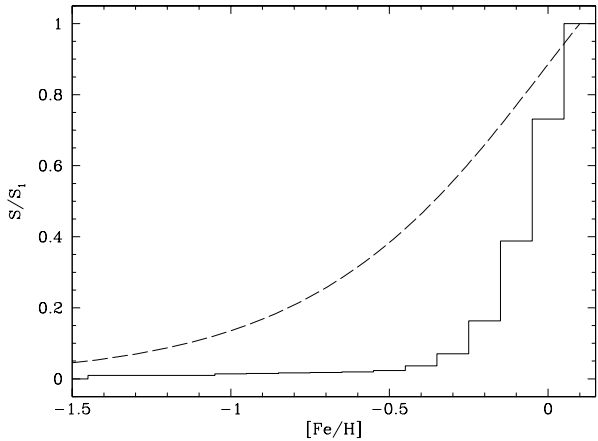, width=6.2cm, angle=270}
\caption{Theoretical (dashed curve) and observed (solid line) cumulative
stellar metallicity distributions.
The theoretical values are calculated using the Simple model. ${S}/{S_1}$
represents the fraction of stars that have metallicites less than $Z$.
}
\label{F3}
\end{figure}

In Figure \ref{F3} we compare $S/S_1$ to the value calculated from our
M dwarf metallicities. We calculated $S/S_1$ from equation \ref{eq:tinsley}
where we assume that $\mu_1= 0.27$ \citep{h00} and
let $\log(Z_1/Z_\odot)=+0.1$.
The theoretical $S/S_1$ calculated for the Simple model
will be somewhat different for other assumed $Z_1$ and $\mu_1$, and a fairly
large range of values for $\mu_1$ are possible, given the 50 percent
uncertainty reported by \cite{h00}.  But for all
reasonable values the theoretical curve drops off much more slowly toward low
metallicity than the observed curve does.

\section{Conclusions}
Our data show that there is an M dwarf problem similar to the previously known
G and K dwarf problems.  The number of M dwarfs peaks at [Fe/H] = 0.0 and drops
off quickly at lower metallicities.

We must be careful in comparing our M dwarf numbers to models of Galactic
chemical evolution because our M dwarfs are not necessarily
representative of the solar neighbourhood.
Galactic chemical evolution models normally include a region about 1 kpc wide
in the Galactic plane and about 1 kpc perpendicular to the plane to include
stars that formed in the thin disc but which now leave the local neighbourhood
\citep{t80}.
The SDSS magnitude limits mean that our sample reaches out to about
2 kpc \citep{a06} and that no nearby M dwarfs are included.
The part of the survey which includes our M dwarf spectra is
centred near the southern Galactic cap \citep{b07a}.
Much of the sampled region is
more distant than the thin disc scale height, so we expect a larger fraction of
thick disc stars to be included in the sample than would be found locally.

Including a larger fraction of stars outside the thin disc would presumably
mean including more low-metallicity stars than would be found in a sample
from the local neighbourhood.  The fact that even with the probable inclusion
of more thick disc stars we still find a paucity of low metallicity stars
strengthens the case for the existence of an M dwarf problem.

In addition to our sample being made up of non-local stars, the fraction of
thin disc versus thick disc stars should vary with metallicity because
metallicity
is correlated with luminosity and distance for late main sequence stars.
We corrected for differences in sampling volume caused by low-metallicity stars
being less luminous and thus being observed at smaller distances for similar
apparent magnitudes.  We did not correct for the difference in population
sampling caused by the variation of brightness with metallicity. Nor did we
correct for differences in average stellar metallicity with 
distance perpendicular to the disc. 
Solar metallicity stars in our sample are at a larger average distance from
the Galactic disc than low-metallicity stars with the same temperature.
We therefore expect the solar metallicity stars to be found in a region with a
larger fraction of thick disc stars and a smaller average metallicity.
Ivezi\'c et al. \citeyearpar{i08} found no correlation
between metallicity and kinematics in the Galactic disc, but did measure a
0.2 dex variation in metallicity from 500 pc to several kpc above the disc.
Because the metallicity variation with distance in our sample should be small,
less than the 0.2 dex variation found by \citet{i08}, our lack of a correction
for metallicity or population differences in our sample should not alter our
results appreciably.

We find evidence that the small fraction of low-metallicity stars seen for
G dwarfs is observed for M dwarfs as well, providing another
strong constraint on models of Galactic chemical evolution.  
The solution to the G, K, and M dwarf problem will require using models which
abandon the assumptions of the `Simple model' as discussed in the
introduction.  

\section*{Acknowledgments}

V.W. gratefully acknowledges the financial support of the Kenilworth Fund of the
New York Community Trust.
We thank George Wallerstein for helpful suggestions on the work and the
manuscript.
Funding for the Sloan Digital Sky Survey (SDSS) and SDSS-II has been
provided by the Alfred P. Sloan Foundation, the Participating Institutions,
the National Science Foundation, the U.S. Department of Energy,
the National Aeronautics and Space Administration, the Japanese
Monbukagakusho, and the Max Planck Society, and the Higher Education Funding
Council for England. The SDSS Web site is http://www.sdss.org/.
The SDSS is managed by the Astrophysical Research Consortium (ARC) for the
Participating Institutions. The Participating Institutions are the American
Museum of Natural History, Astrophysical Institute Potsdam, University of
Basel, University of Cambridge, Case Western Reserve University, The
University of Chicago, Drexel University, Fermilab, the Institute for Advanced
Study, the Japan Participation Group, The Johns Hopkins University, the Joint
Institute for Nuclear Astrophysics, the Kavli Institute for Particle
Astrophysics and Cosmology, the Korean Scientist Group, the Chinese Academy of
Sciences (LAMOST), Los Alamos National Laboratory, the Max-Planck-Institute for
Astronomy (MPIA), the Max-Planck-Institute for Astrophysics (MPA), New Mexico
State University, Ohio State University, University of Pittsburgh, University of
Portsmouth, Princeton University, the United States Naval Observatory, and the
University of Washington.

\end{document}